# Tailoring Mechanical and Acoustic Properties of Liquid-Filled Elastomers as Reusable Ultrasound Couplants

Surojit Ranoo[1*], Romain Fayolle[2], Jean Baudry[1], Nicolas Bremond[1*]

1. LCMD, CBI, ESPCI Paris, Université PSL, CNRS 75005 Paris, France
2. ID4US, 38400 Saint Martin d'Heres, France

Correspondence: Surojit Ranoo (surojit.ranoo@espci.fr) and Nicolas Bremond (nicolas.bremond@espci.fr)

**Abstract**

Couplants are indispensable for ultrasound-based applications to ensure seamless propagation of the acoustic waves between the transducer and targets. Existing hydro/oil-based gel couplants provide excellent acoustic coupling but are single-use and therefore not suitable for emerging applications such as wearable health monitoring, ultrasound-based biometric devices, and prolonged sonodynamic therapy. Alternatively, polymer-based solid couplants have been explored as reusable solutions but generally fail to simultaneously provide effective mechanical conformability and efficient acoustic transmission. Here, we propose an innovative composite material design based on a silicone elastomer matrix, in which mechanical and acoustic properties are tailored independently through the microencapsulation of diols and triols. An emulsion-based material formulation strategy is employed to produce a family of high-liquid droplet content elastomeric composites containing up to 75 volume fraction (%) of liquid. A robust material formulation strategy is established that enables independent parametric control over the resulting mechanical and acoustic properties. A specific composition containing 70 % glycerol has been identified and explored for biomedical applications, exhibiting mechanical softness comparable to that of skin and acoustic properties matching those of soft tissues. The material has been integrated into a functional ultrasound characterization device, demonstrating structural homogeneity and applicability of the proposed couplant. Furthermore, the composites exhibit mechanical conformability to stiff substrates with surface irregularities, making them also promising candidates for industrial applications.

## 1. Introduction

Ultrasound-based technologies have become indispensable across biomedical diagnostics and industrial non-destructive testing (NDT), enabling rapid imaging, real-time 3D visualization, structural interrogation, and assessment of material integrity.[1][2][3] In clinical settings, ultrasound supports biological tissue imaging, cardiovascular assessment, and functional organ monitoring, while its portability and excellent safety profile have accelerated the development of point-of-care diagnostics and continuous physiological monitoring.[4][5] More recently, advances in flexible transducers, wearable ultrasound devices, phased array systems, and high-intensity focused ultrasound technology have expanded ultrasound beyond conventional imaging toward continuous health monitoring, secure biometrics, targeted therapy, and deep tissue modulation, placing new demands on coupling materials capable of prolonged or repeated operation.[6][7][8][9]

Efficient acoustic coupling is essential for all ultrasound-based technologies. At operating frequencies typically ranging from 1 to 25 MHz, even microscopic air gaps at the interface cause severe acoustic reflections due to the large impedance mismatch, substantially degrading signal transmission and imaging performance. Consequently, coupling materials remain indispensable despite remarkable advances in ultrasound transducers and signal-processing techniques. However, the development of coupling materials has progressed considerably slower than advances in ultrasound hardware, creating an important materials challenge for next-generation ultrasound technologies.

Existing ultrasound couplants for bio-applications are broadly categorized into hydrogel-based couplants, dry polymeric couplants, and semi-dry membrane systems. Hydrogel-based couplants remain the clinical standard because they provide excellent acoustic transmission, lubrication, and ease of use. These are inherently single-use, gradually dehydrate during prolonged use, and are not suitable for wearable devices. Reusable dry couplants based on polyurethane and silicone elastomers have therefore attracted considerable interest. Polyurethane elastomers exhibit favorable acoustic properties but remain relatively stiff and degrade upon exposure to UV and temperatures.[10][11][12] In contrast, silicone elastomers provide biocompatibility, environmental stability, and straightforward fabrication but differ significantly in acoustic properties from those of soft tissues.[13] Semi-dry membrane couplants encapsulating hydrogels or water improve mechanical conformability but require careful thickness ($\leq \lambda/4$) optimization to reduce acoustic mismatch and signal-to-noise ratio.[6][10] These couplants often suffer from dehydration, shrinkage, and wrinkled surfaces, leading to further signal distortion and necessitating additional complex fabrication and optimization steps.[6][14] Consequently, these couplants have not been widely adopted for reusable ultrasound applications. Innovative material designs are proposed utilizing stress-sensitive solid-to-liquid transitioning polysilicone gel and liquid metal included silicone elastomer, but these materials are highly attenuative and acoustically far from soft tissue.[15][16][17]

The lack of suitable coupling materials has diverted efforts toward device-specific solutions in the last decade, particularly for wearable ultrasound systems. Flexible capacitive micromachined ultrasound transducer (CMUT) arrays are designed with innovative architecture and laminated in a thin silicone elastomer layer.[7][18][19][20] These devices are flexible and have demonstrated promising performance in specific device architectures that require complex fabrication processes. These developments have significantly advanced wearable ultrasound technologies and highlight the importance of sensor-form-factor-independent, mechanically compliant, and acoustically matched couplants. However, existing approaches primarily optimize either mechanical compliance or acoustic performance. Consequently, the development of reusable ultrasound couplants remains largely empirical.

From a materials design perspective, the key challenge is not simply developing another ultrasound couplant, but also establishing a rational design strategy capable of simultaneously achieving acoustic matching and mechanical compliance, while enabling their independent control, so it can be adopted for a wide range of sensors and applications (**figure 1g).** In conventional polymeric composites, improving acoustic properties with solid fillers often compromises mechanical softness, whereas reducing crosslink density to enhance conformability typically increases viscoelastic attenuation. Therefore, achieving an optimal balance between acoustic transmission and mechanical compliance has remained a persistent challenge in materials engineering over the past two decades. The requirements suggest that heterogeneous elastomeric composites incorporating liquid droplets provide an attractive materials platform, as the dispersed liquid phase offers a direct means to modify effective mechanical and acoustic responses of the composite; however, the simultaneous optimization of properties has yet to be explored.[21][22] Establishing quantitative relationships between formulation and processing parameters, microstructure, and mechano-acoustic performance is therefore essential for rational materials design.

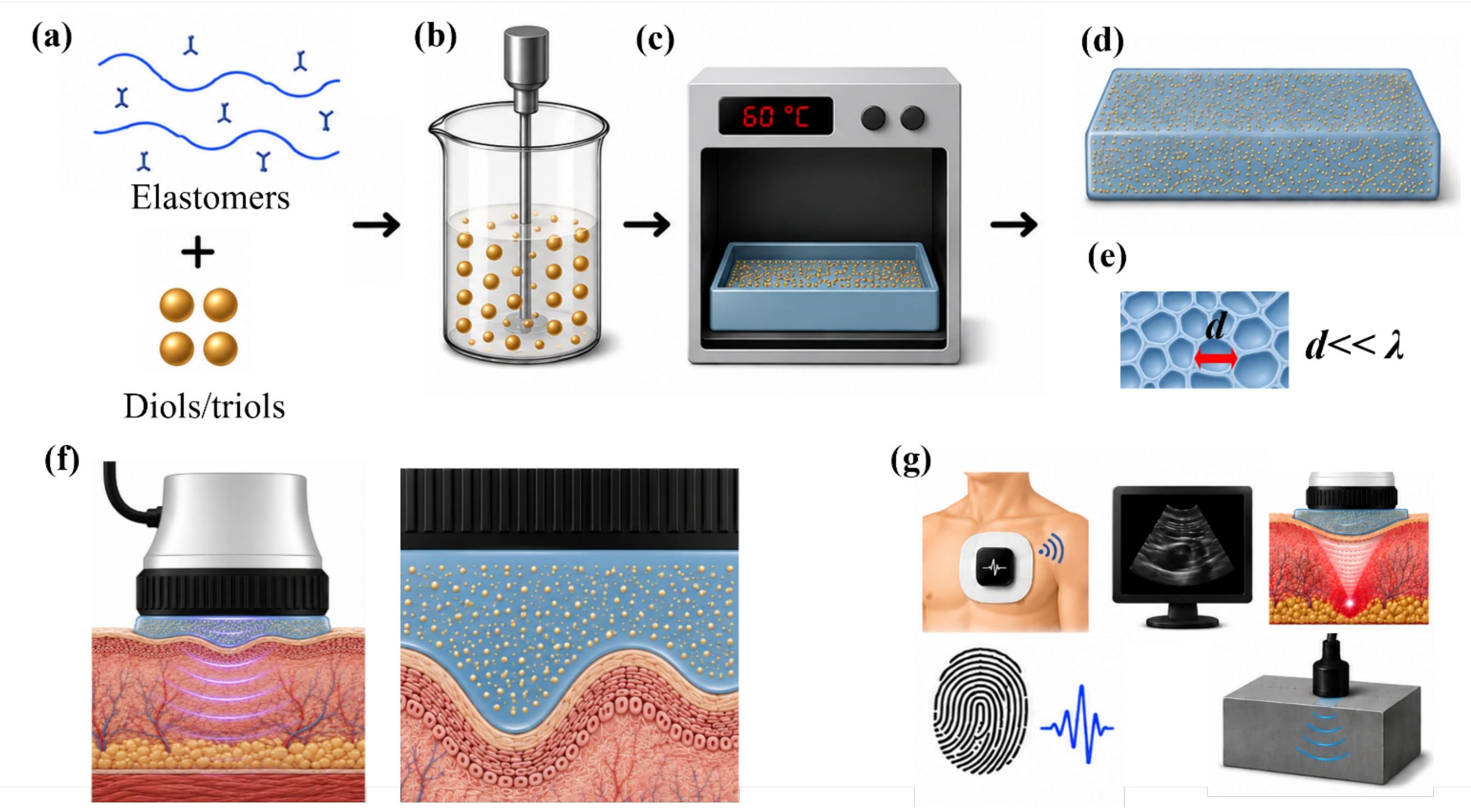

**Figure 1**. Schematic representation of concept, formulation strategy, desired properties, and related applications. (a) Constituents: Continuous phase (elastomer) and dispersed phase (diols/triols). (b) Emulsion preparation under a controlled rate of dispersed phase addition at constant temperature. (c) Curing of the viscoelastic emulsion into a solid elastomeric pad under controlled heating. (d) Soft elastomeric pad with encapsulated liquid droplets. (e) Closed-cell microstructure with cell dimensions significantly less than the ultrasound wavelength. (f) Expected efficient mechanical conformation between the couplant and skin, and transmission of acoustic energy. (g) Highlighting possible applications in wearable ultrasound devices, static ultrasonography, ultrasound therapy, ultrasound-based biometrics, and possible extension to NDT.

Following these considerations, we present a materials design strategy for engineering reusable ultrasound coupling materials based on heterogeneous elastomeric composites with liquid droplet inclusions. A surfactant-free emulsion preparation approach is employed using silicone elastomer as the continuous phase and selected polar liquids as the dispersed phase. Through systematic investigation of formulation, processing conditions, and microstructure, we establish the structure-property relationships governing the acoustic and mechanical behavior of these composites and demonstrate that acoustic and mechanical responses can be independently tailored within a single materials platform. Finally, we demonstrate the versatility of the proposed strategy by developing a reusable coupling material with soft-tissue-like acoustic properties and skin-like mechanical compliance, and by successfully integrating it into a portable imaging device. Beyond this formulation, the proposed framework provides a versatile foundation to engineer application-specific ultrasound couplant for biomedical and industrial applications.

## 2. Results and Discussion

In this article, we systematically describe a material design strategy to develop a family of reusable ultrasound couplants. Guided by the materials design strategy outlined in **figure 1**, the Results and Discussion are organized into five parts: (i) theoretical framework, (ii) materials formulation, (iii) structure-property relationship, and (iv) validation of selected composition for applications.

### 2.1 Theoretical framework

For a heterogeneous elastomeric composite containing liquid droplets, the elastic modulus is governed by the liquid volume fraction ($\varphi$), and microstructure. For low $\varphi$, the system is an elastomer containing discrete liquid droplets. The effective Young's modulus of such liquid-filled elastomer composites is described using classical linear elasticity theory based on Eshelby's model[21][23]

$$E_c = [1/(1 + \frac{5}{3}\varphi)]E_{SE} \quad (1)$$

Here, $E_c$ and $E_{SE}$ are the Young's moduli of the elastomeric composite and the silicone elastomer, respectively, $\varphi$ is the liquid volume fraction. Equation 1 provides an effective route to achieving a mechanically compliant elastomeric system by increasing $\varphi$. However, as $\varphi$ exceeds 60 %, the droplets become closely packed and approach the random close-packing limit, leading to a gradual transition to a foam-like microstructure. The mechanical response of such cellular solids is commonly described by the Gibson-Ashby model.[24][25] However, for closed-cell foams containing incompressible liquid droplets, deformation is also influenced by the incompressibility of the liquid phase. Warner *et al.* proposed a two-component constitutive model describing the elastic response of liquid-filled closed-cell foams:[26]

$$E_c = [C_1(1 - \varphi) + C_2\,(1 - \varphi)^2]E_{SE} \quad (2)$$

Here, $\varphi$ indicates the liquid volume fraction. $C_1$ and $C_2$ are constant ($C_1 + C_2 = 1$). The second term represents bending of the cell walls, which is analogous to the mechanical behavior of conventional dry foams. The first term accounts for the stretching of the cell walls, which arises because the incompressible liquid is trapped within each closed cell. Consequently, deformation initially proceeds through cell-wall bending, followed by stretching of the cell walls to accommodate the incompressible liquid.[26] Equation 1 and 2 together suggest an overall decrease in $E_c$ with increasing $\varphi$. Commercially available low-attenuative silicone elastomers typically possess hardness exceeding 50 Shore A ($E_{SE} \geq 2.5$ $MPa$). Achieving skin-like softness (20-50 Shore 00, 0.2-0.4 $MPa$) requires the incorporation of liquid droplets above the random close-packing limit in silicone elastomer.

In addition to mechanical compliance, the acoustic behavior of liquid-filled elastomer composites must be carefully considered. At every solid-liquid interface, acoustic waves undergo reflection and refraction due to acoustic impedance mismatch. In addition, a large number of liquid droplets contribute to multiple scattering of the wave. These effects result in very high acoustic attenuation. This detrimental effect can be substantially reduced by minimizing the droplet size, ideally to less than 5 % of the wavelength. Under such conditions, scattering becomes negligible, and the interfacial effects are effectively suppressed. The resultant speed of sound for such a composite can be described using Wood's effective medium theory as[22][27][28]

$$v_c = \sqrt{K_c/\rho_c} = \sqrt{1/[\varphi\rho_l + (1-\varphi)\rho_s] \times [\frac{\varphi}{\rho_l v_l^2} + \frac{(1-\varphi)}{\rho_s v_s^2}]} \qquad 3$$

where $v_c$ is the effective sound speed in the elastomer composites and $\varphi$ represents the volume fraction of liquid inclusion, $K_c$ and $\rho_c$ are the effective bulk modulus and effective density of the composites, respectively, defined as $1/K_c = \varphi/K_l + (1-\varphi)/K_s$ and $\rho_c = \varphi\rho_l + (1-\varphi)\rho_s$. $K_s$ and $\rho_s$ are the bulk modulus and density of the solid elastomer, and $K_l$ and $\rho_l$ are the bulk modulus and density of the liquid. For solid-liquid composite, this theory is applicable only to solids with a very low shear modulus. The sound speed of liquid-filled elastomer composites is a volumetric property governed primarily by the total liquid volume and the intrinsic sound speed of the liquid. Excluding liquid metals, only a limited number of liquids have sound speeds exceeding that of soft tissue, considerably restricting the choice of suitable dispersed liquids.

### 2.2 Formulation of the composite materials

For ease of formulation, a commercially available silicone elastomer (SE), Sylgard 184, was selected as the continuous phase. It is non-toxic, thermally stable up to 150 °C, and widely used for biomedical applications.[13][29] The intrinsic speed of sound of cured Sylgard 184 (1000-1050 m $s^{-1}$) is substantially lower than that of soft tissues (~1500 m $s^{-1}$), necessitating the incorporation of liquids with sound speed exceeding 1500 m $s^{-1}$ to achieve acoustic matching.[30][31] The liquids satisfying these criteria are listed in Table 1. According to equation 3, the required liquid volume fraction is ≥ 75 %. Although glycerol-in-SE emulsions have been reported for $\varphi < 50$ % using surfactants, preparing high-internal-phase emulsions ($\varphi > 70$ %) without surfactants and with a very small droplet size is considerably more challenging.[32][33]

Recently, Nannette *et al.* demonstrated stable emulsions containing up to 80 vol. % dispersed phase in viscous silicone oil through matching viscosities of silicone oil and aqueous phases.[34] The silicone elastomer consists of two components with a recommended mixing ratio of 10:1 (A:B), which exhibits a viscosity of ~ 3.5 Pa s. Emulsions prepared using 10:1 ratio with various dispersed phases exhibited pronounced viscoelasticity above $\varphi = 45\,\%$. The increasing viscoelasticity restricts the incorporation of additional liquid beyond a maximum $\varphi$ (denoted as $\varphi^*$). Further addition of liquid resulted in gradual phase inversion. Among all the liquids, glycerol exhibited the highest attainable $\varphi^*$ of 56%. Adjusting the composition to 10:2 reduced the viscosity to 1.92 Pa s and resulted in increased $\varphi^*$ to 63%, which remained well below the target. Further increasing the Part B proportion was avoided as it increases crosslink density, producing a stiffer elastomer.

In the present system, the viscosity of the sylgard 184 (10:2) is twice of glycerol and an order of magnitude higher than that of the diols. To improve viscosity matching, low viscous silicone oils with different PDMS chain lengths and functionalities were incorporated into the sylgard 184. The optimum formulation contained 30 vol. % silicone oil, resulting in a continuous-phase viscosity of 1.1-1.2 Pa s (Table 1). This modified elastomer enabled $\varphi^*$ values above 70 % for glycerol, whereas $\varphi^*$ for EG remained $< 64\,\%$ and also for DEG and TEG. Therefore, $\varphi = 50\,\%$ was considered to investigate the influence of the dispersed liquids. After formulation, the emulsions were subsequently cast into customized molds and cured under controlled conditions (Section 4.3). A schematic representation of the fabrication process is shown in **figure 1**.

**Table 1.** Physical properties of the continuous and dispersed phases

| Continuous phase (Elastomer matrix) | | | | Dispersed phase (Liquid droplet) | | | |
|---|---|---|---|---|---|---|---|
| Code | Composition | Viscosity Pa s | Crosslink density per $m^{-3}$ | Code | Composition | Viscosity Pa s | Sound speed m $s^{-1}$ |
| **S1** | Sylgard 184 | 1.92 | $5.53\times10^{26}$ | **Gly** | Glycerol | 0.890 | 1904 |
| **S2** | Sylgard 184 + vinyl-terminated silicone oil (500 cSt) | 1.22 | $4.20\times10^{26}$ | **EG** | Ethylene glycol | 0.018 | 1658 |
| **S3** | Sylgard 184 + silicone oil (350 cSt) | 1.12 | $2.09\times10^{26}$ | **DEG** | Diethylene glycol | 0.034 | 1586 |
| **S4** | Sylgard 184 + silicone oil (500 cSt) | 1.18 | $1.76\times10^{26}$ | **TEG** | Triethylene glycol | 0.047 | 1608 |

To investigate the influence of the dispersed liquid, emulsions with $\varphi = 50\%$ were prepared using silicone oil-modified elastomer (S3). The as-prepared emulsions were characterized using optical microscopy. The EG droplets were significantly larger than those obtained with glycerol under identical processing conditions, as shown in **figure 2a** and **2b**. The droplet size for DEG and TEG was even larger, as shown in **figure S7**, indicating progressive droplet coalescence and instability. This behavior is reflected in the microstructure of the cured composites. Composites prepared with glycerol exhibited a uniform closed-cell microstructure containing relatively small pores, whereas those prepared with EG exhibited an interconnected open-cell structure.

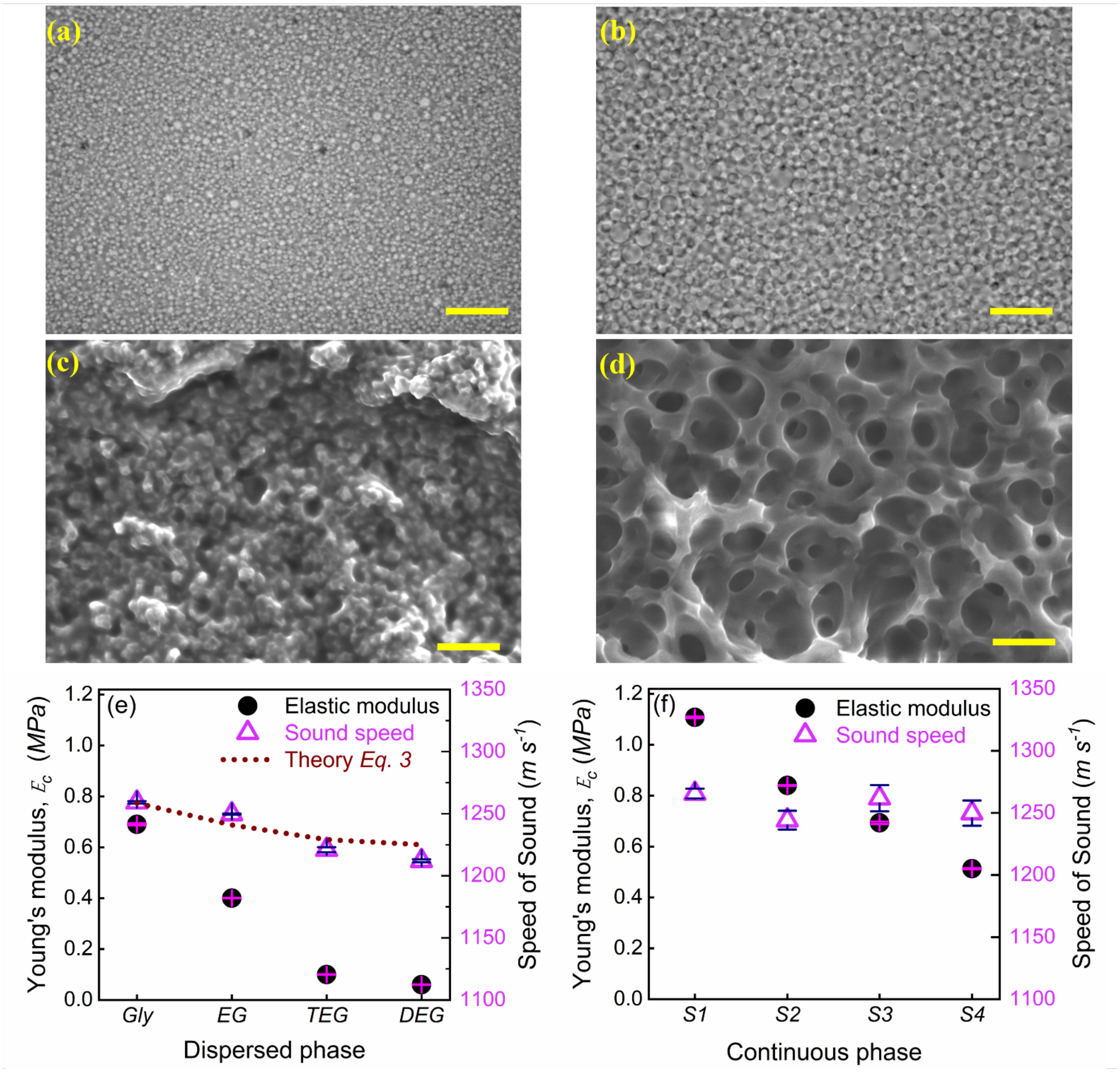

**Figure 2**: Optical microscopy images of as-prepared emulsion with (a) glycerol and (b) EG of 50 volume fraction. Electron microscopy images of the microstructure of solid composites incorporated with (c) glycerol and (d) EG of 50 volume fractions. Variation of materials properties of solid composites formulated with (e) S3 and 50 % of different dispersed (liquid) phase and (f) different elastomer compositions and 50% glycerol. The scale bar corresponds to 10 $\mu m$.

**Figure 2e** presents the Young's modulus ($E_c$) and sound speed of composites with different dispersed liquids. $E_c$ exhibits substantial variation, which originates from the distinct microstructures of the composites formulated with different dispersed phases. The DEG and TEG-based formulations exhibited very low $E_c$, gel-like behavior, and did not form mechanically stable elastomeric pads. This behavior originates from the reaction of the hydroxyl (-OH) groups of the long-chain glycols with the Si-H groups of the methylhydrosiloxane-dimethylsiloxane copolymer through a platinum-catalyzed hydrosilylation reaction, forming amphiphilic copolymers at the droplet interface.[35][36] These copolymers reduce emulsion stability and produce inhomogeneous crosslinking within the elastomer matrix. The underlying reaction mechanism is discussed in the Supplementary Information. The sound speed of all formulations follows the prediction of Equation 3 regardless of droplet size or microstructure. Since open-cell or gel-like structures are not suitable for reusable couplants, glycerol was selected as the dispersed phase for all subsequent formulations.

The influence of the continuous phase was subsequently investigated using glycerol-filled composites prepared with different elastomer matrices. As shown in **figure 2f**, the $E_c$ decreased 3 times for different elastomer compositions, demonstrating that, for constant φ, the mechanical response is governed primarily by the elastomer network. For an ideal elastomer network, the Young's modulus is related to the crosslink density (ν) through $E_{SE} = \frac{3}{2}\nu k_B T$, where $k_B$ is the Boltzmann constant, and $T$ is the temperature.[37][38] The observed variation in the effective $E_c$, therefore, originates primarily from changes in the crosslink density of the elastomer matrix. In contrast, the sound speed remained nearly constant across all formulations. This behavior arises because ultrasound propagation is governed mainly by the bulk modulus of the constituent phases, which is only weakly affected by changes in crosslink density. Consequently, the acoustic response remained essentially similar despite the substantial variation in mechanical stiffness. These observations demonstrate that the mechanical compliance of the elastomeric couplant can be tailored independently by modifying the elastomer network, while preserving its acoustic performance.

To achieve the liquid volume fraction required for acoustic matching, the effect of $\varphi$ was investigated in detail using glycerol in the silicone oil-modified silicone elastomer (S3). Stable emulsions containing up to 75 vol. % glycerol were successfully prepared. The stable emulsions were characterized by optical microscopy, as shown in **figure S2**. The stability is attributed to the formation of a highly viscous PDMS film among the droplets due to the reorganization of PDMS molecules at the droplet interface, which remained even after a 20-fold dilution (**figure S3**). The droplets were strongly adhesive due to this thin film. The freshly prepared emulsions exhibited pronounced viscoelasticity, characterized by shear-thinning and yield-stress behavior. For $\varphi = 70$ %, the low-shear viscosity at 0.01 $s^{-1}$ was approximately $5.8 \times 10^5$ Pa s, decreasing to approximately 47 Pa s at a shear rate of 1 $s^{-1}$. The yield stress increased from approximately 30 to 500 Pa as $\varphi$ increased from 50 to 70%, as shown in **figure S4**. For $\varphi \geq 70\%$, the droplet size increased markedly because the increasing viscoelasticity hindered efficient mixing. These emulsions were subsequently homogenized manually using a mortar and pestle. The glycerol droplet diameter ranged from approximately 1 to 3 μm, corresponding to less than 5% of the ultrasound wavelength at 10 MHz (assuming a sound speed of 1300-1500 m $s^{-1}$). According to the theoretical criterion established in Section 2.1, droplets of this size are expected to produce negligible ultrasound scattering while maintaining high acoustic transparency.[39]

### 2.3 Structure-property relationship

#### 2.3.1 Mechanical properties

The mechanical properties of the elastomeric pads were evaluated under uniaxial compression. The experimental procedure is described in detail in Section 4.6. The reproducibility of the measurements was verified through multiple consecutive compressions, different sample batches, and measurements performed after several weeks of storage (Section S3, **figure S5,** and **figure S6**, in the supplementary information).

The variation of Young's modulus ($E_c$) with increasing glycerol volume fraction is presented in **figure 3a**. Overall, the elastic modulus decreases progressively with increasing $\varphi$, with a distinct transition occurring approximately at 60 % of glycerol. At relatively low liquid droplet content, the reduction in $E$ primarily arises from the gradual replacement of the solid elastomer by mechanically compliant liquid droplets. For liquid-filled elastomeric composites composed of an elastomer with a high elastic modulus (in MPa) the interfacial surface energy is negligible. The mechanical behavior is described by the classical linear elasticity model, and accordingly governed by equation 1.[40]

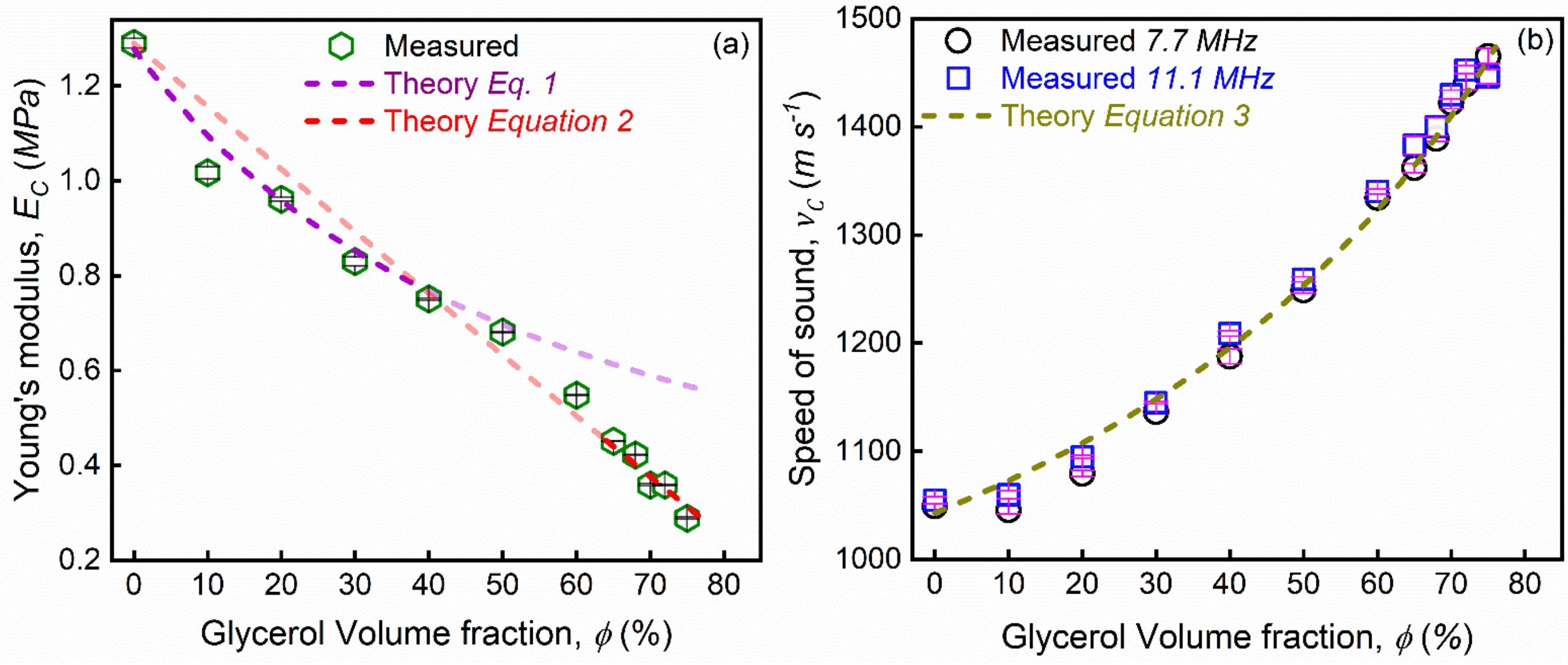

**Figure 3:** (a) Variation of the Young's modulus ($E_c$) of elastomeric composites with the inclusion of glycerol volume fractions. Theoretical model fittings describing the microstructural transformation. (b) Variation of the speed of sound of elastomeric composites with glycerol volume fractions ($\varphi$) and theoretical fitting.

Equation 1 accurately reproduces the experimental data up to 40 % glycerol but progressively overestimates the modulus at higher liquid droplet contents. As $\varphi$ exceeds 60 %, the droplets become closely packed and approach the random close-packing limit, leading to a gradual transition to a foam-like microstructure. For closed-cell foams containing an incompressible liquid, deformation is additionally influenced by the incompressibility of the liquid phase and defined by equation 2. This model is applicable to low solid volume fractions (approximately 0.1-0.3), corresponding to the $\varphi \geq 70$ % in this work. Accordingly, it accurately captures the experimentally observed behavior of the developed composites ( $\geq 70$ %). However, a transition towards foam-like mechanical behavior is already evident at $\varphi \geq 65$ % although the corresponding solid volume fraction > 0.3. The strong adhesion among the droplets promotes the formation of localized pockets of foam-like microstructure within the elastomer matrix. These contributed to a foam-like mechanical response even at relatively higher solid fractions. Overall, for a particular composite, the mechanical response is primarily governed by $\varphi$, which influences the evolution of microstructure.

**2.3.2 Acoustic properties**

For acoustic characterization, rectangular composite pads were prepared as shown in **figure S8**. The acoustic properties of the elastomeric pads were characterized using a custom-built transmission-mode experimental setup (Section 4.5), with deionized water serving as the reference medium. Measurements were performed over a frequency range of 7.7-14.3 MHz. The sound speed was determined from the time delay of the transmitted ultrasound signal introduced by the elastomeric pad placed between two transducers. A pronounced increase in the sound speed was observed with increasing glycerol volume fraction, as shown in **figure 3b**. At very high $\varphi$ ( $\geq 70$), the sound speed approached that of the soft tissue.

The observed trend can be accurately described using the effective medium approximation, as in Equation 3, proposed by Wood for non-interacting biphasic systems[22],[16],[27],[28]. The good fitting ensures a non-interacting system and minimal incorporation of air bubbles. These results, together with the SEM observations, confirm that the fabricated pads contain negligible air inclusions, which is a critical requirement for ultrasound imaging applications. The contribution of the shear modulus of the elastomer matrix is neglected in the effective medium approximation. In general, solids possess a finite shear modulus, and the longitudinal sound speed ($\upsilon_L$) is governed by the longitudinal elastic modulus ($M_L$) according to $\upsilon_L = \sqrt{M_L/\rho}$ , where $M_L = K_s + 4/3\, G_s$ and $K_s$ and $G_s$ denote the bulk and shear moduli, respectively.[41] However, for PDMS-based silicone elastomers, the shear modulus ($G_s \sim MPa$) is approximately three orders of magnitude smaller than the bulk modulus ($K_s \sim GPa$). Consequently, its contribution to the longitudinal modulus is negligible, and Wood's effective medium model is applicable to the present composite system.[41] The acoustic impedance of composites, calculated from the density and speed of sound, also increases with $\varphi$ and spans the impedance range of biological tissues, including fat, skin, blood, and muscle, thereby enabling efficient acoustic matching (**figure S9**).[42] These observations confirm that the sound speed and acoustic impedance effectively depend on the volumetric property of the liquid and solid phases. These results established that liquid inclusion is a primary design parameter governing acoustic transmission, thereby enabling straightforward tuning of the sound speed through appropriate selection of the liquid and its volume fractions.

The combined acoustic and mechanical properties presented in **figure 3** demonstrate a remarkable dual effect of increasing liquid content ($\varphi$). Increasing $\varphi$ simultaneously increases the sound speed, thereby improving acoustic coupling, while reducing the Young's modulus, thereby enhancing mechanical conformability. The mechanical properties are a function of the elastomer network and microstructure, whereas the acoustic properties are volumetric and governed by the total liquid droplet volume. The formulation is robust and scalable, the processing is simple, and the material can be fabricated into desired shapes and sizes.

### 2.4 Validation of selected composition for applications

Overall, the glycerol-silicone elastomer exhibits the most favorable combination of microstructural stability, acoustic performance, and mechanical properties. Composite pads containing 70 % glycerol possess an acoustic impedance of ~1.66 MRay*l* and a hardness of 45 Shore 00, closely matching the acoustic impedance, sound speed, and mechanical characteristics of human skin.[43] These excellent acoustic impedance matching, together with sufficient mechanical compliance, make them suitable for biomedical applications in humans and animals. The attenuation for this composite is 2.64 dB $mm^{-1}$ at 11.1 MHz.

The couplant was wiped with ethanol without any measurable change in its properties. The developed materials exhibited long-term stability under laboratory conditions (24 ± 3 ℃), with only a 4.8 % change in mass and a 1% change in speed of sound after 6 months of storage. They also demonstrated good stability under controlled environmental conditions of 30 ± 0.1 ℃ and 30% relative humidity for 15 days, followed by 30 ± 0.1 ℃ and 40% relative humidity for 7 days. Under these conditions, the pad exhibited a change in mass of approximately 1.4 % and a 2.1 % decrease in sound speed after 21 days. This performance represents a substantial improvement over previously reported gel-filled membrane couplants and highlights the suitability of the developed materials for ultrasound devices intended for environmental exposure.

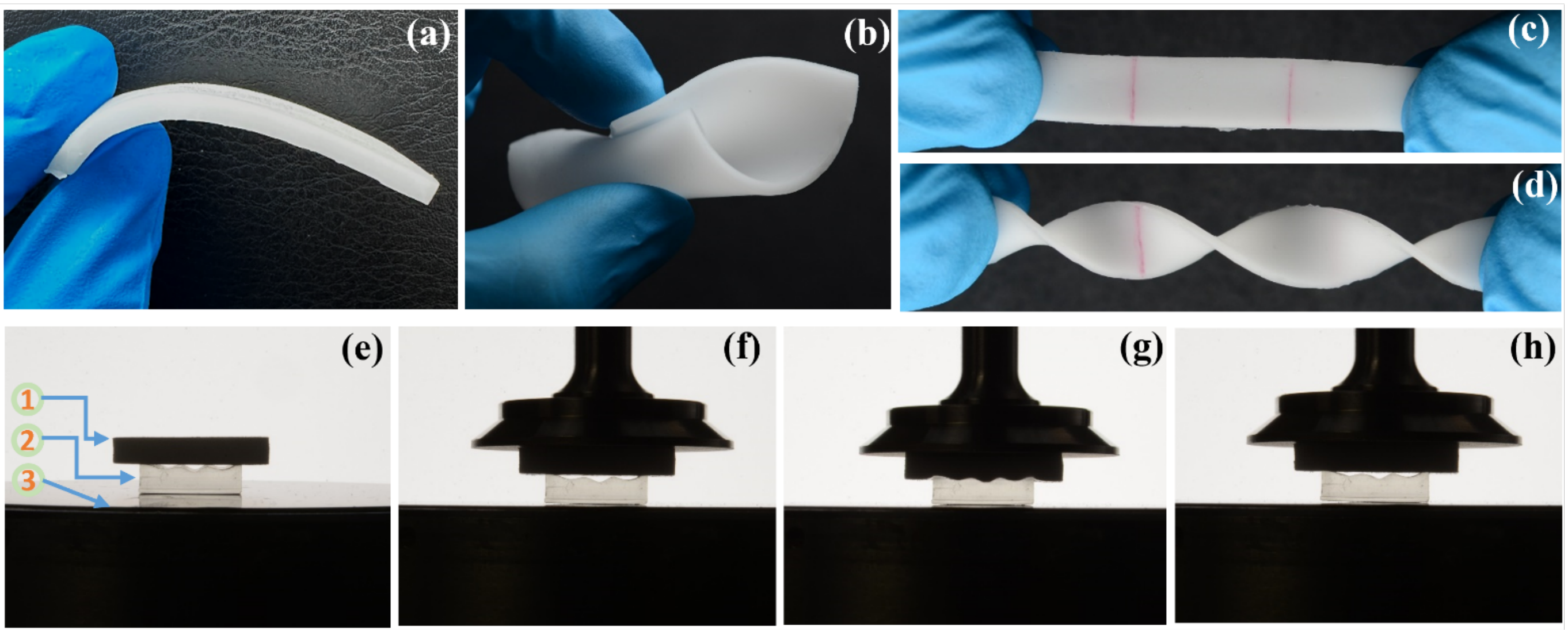

**Figure 4**: Photographs displaying a compliant soft pad capable of (a) bending (4 mm thick sample*)*, (b) folding (2 mm thick sample*)*, (c) and (d) stretching (100 %) and twisting (1.2 *mm* thick sample). Mechanical compliance test experiment illustration (e) with (1) developed couplant with 70 % glycerol, (2) 3D printed stiff materials with wavy topographic feature, and (3) bottom plate of rheometer. Progressing compression test snapshots (f) at the beginning, (g) fully conformed state, and (h) post compression.

#### 2.4.1 Mechanical conformability

The mechanical conformability of the selected elastomeric couplant is qualitatively shown in **figure 4a-d** including bending, folding, stretching, and twisting. For this soft couplant, mechanical conformability to human skin is comparatively easy to achieve due to the inherent softness of the skin. To provide a more rigorous assessment of the developed material, its mechanical conformability is evaluated using a resin-based 3D-printed stiff substrate possessing well-defined wavy surface topographies, which represent a considerably demanding contact condition. The experimental setup is described in section 4.7. The elastomeric pad was positioned between the patterned substrate and the upper compression plate (flat rheometer geometry). Upon application of axial load, the initially flat lower surface of the elastomeric couplant conforms progressively to the wavy surface profile of the rigid substrate, as illustrated in **figure 4e-h**. Complete conformal contact was achieved without visible air gaps. Upon removal of the compressive load, the pad recovered its original shape, demonstrating excellent reversible mechanical compliance. These results highlight the ability of the material to establish intimate contact with uneven rigid surfaces, thereby extending its potential applications beyond biomedical ultrasound to industrial non-destructive testing (NDT).

### 2.4.2 Acoustic transmission

To demonstrate the practical applicability of the developed coupling material, a fully functional prototype is fabricated and interfaced to a portable ultrasound imaging and characterization device (EcrinLab) developed by ID4US. For the present study, the imaging module consisted of an array of capacitive micromachined ultrasound transducer (CMUT) composite elements integrated with the developed elastomeric couplant. The uncured couplant was deposited onto the CMUT sensor by pressure-assisted extrusion and subsequently cured *in situ* under carefully controlled thermal conditions. Following curing, the elastomeric couplants adhered strongly to the sensor surface without requiring any additional adhesive. For static scenes, the device supports B-mode imaging, full 3D volume reconstruction, and surface imaging, thereby enabling a comprehensive evaluation of the developed coupling material. An image of the device is shown in **figure 5a,** and more information is provided in section 4.8.

The imaging performance was evaluated using a custom-designed agar-based tissue-mimicking phantom containing embedded nylon wire targets (described in Section 4.8). Representative B-mode images obtained using the developed couplant and reference commercial Aquaflex gel pad (Parker Laboratories) of comparable thickness ($4.5 \pm 0.2$ *mm*) are shown in **Figure 5c** and **5d**. The wire target located up to a depth of 23 mm is clearly resolved in both cases. Encouragingly, the developed couplant provides comparable imaging performance to a commercial disposable aqueous gel pad. Furthermore, in combination with a thinner free-standing couplant (2.2 mm), the EcrinLab system successfully resolved targets at depths of up to 40 mm. **Figure 5e** and **5f** present the reconstructed 3D image of the phantom using the developed elastomeric couplant and commercial pad, respectively. The reconstructed volume exhibits structural uniformity without any detectable artifacts, distortions, or local variations in sharpness or contrast of the scene, demonstrating the homogeneous nature of the developed coupling material and its excellent conformability and acoustic matching to the tissue-mimicking phantom. Compared with commercial couplant of similar thickness, the developed material enables unraveling of finer details of deeper targets, confirming its ability to provide efficient ultrasound transmission. These results demonstrate the suitability of the developed material for prolonged usage in wearable ultrasounds and ultrasound-based biometric devices.

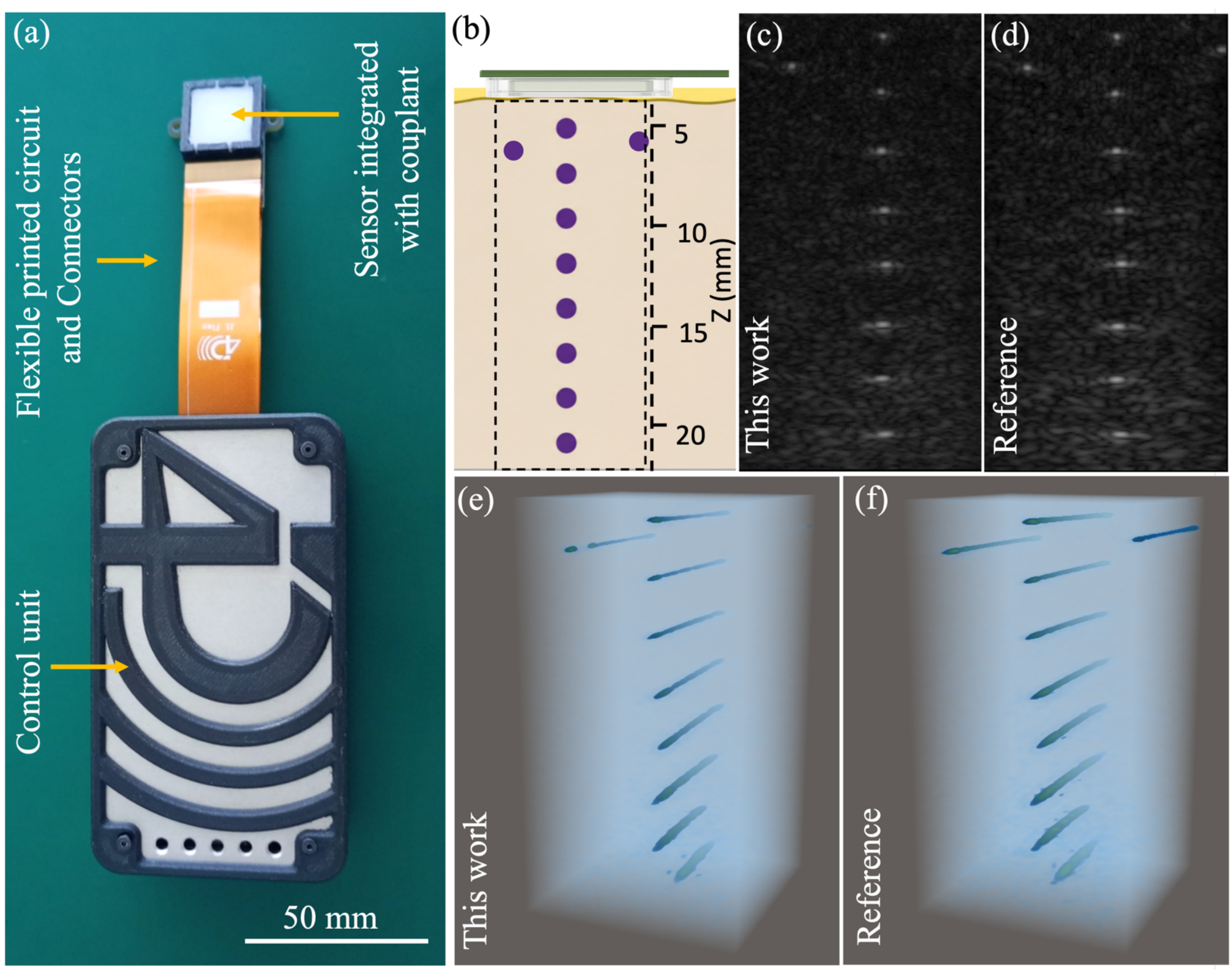

**Figure 5**: (a)Photograph of functional portable ultrasound characterization device Ecrinlab with integrated elastomeric couplants. (b) Schematic of the ultrasound imaging experiment with a tissue-mimicking phantom using Ecrinlab. B-mode image using (c) elastomeric couplant and (d) commercial hydrogel pad couplant. Full matrix capture imaging; 3D reconstruction of tissue-mimicking phantom using (e) elastomeric couplant and (f) commercial hydrogel pad couplant.

Furthermore, the couplant is independent of the sensor architecture; it can be readily integrated with a wide variety of ultrasound transducer configurations and fabricated with application-specific thicknesses to satisfy the requirements of different ultrasound devices. In addition, the uncured material exhibits intrinsic adhesiveness to the silicon-based passivation layer on top of the sensor, enabling direct bonding to ultrasound sensors without requiring a separate adhesive layer, thereby simplifying device integration. These can be adopted for conventional medical probes, which commonly use a silicon-based passivation layer.

To further demonstrate its versatility beyond biomedical imaging, the mechanical conformability of the developed couplant was evaluated using a resin-based 3D-printed stiff substrate possessing well-defined wavy surface topographies (pitch ~ 1 mm). Upon compression, the elastomer conformed to the surface topography, demonstrating conformal contact with the patterned surface. These observations further highlight the potential of the developed coupling material for applications extending beyond biomedical ultrasound, including industrial non-destructive testing.

## 3. Conclusion

A new family of coupling materials, composed of microencapsulated liquid droplets within an elastomer matrix for ultrasound-based applications, has been developed via emulsification. A formulation strategy has been demonstrated to simultaneously tailor two key properties required for efficient ultrasound coupling, namely the sound speed (acoustic impedance) and the Young's modulus. The respective roles of the silicone elastomer (continuous phase) and the liquid inclusions (dispersed phase) in governing emulsion stability, microstructural evolution, and the resulting macroscopic acoustic and mechanical properties have been systematically investigated. The processing conditions were carefully optimized to maintain liquid droplet diameters below 5% of the ultrasound wavelength commonly employed in biomedical imaging, thereby minimizing ultrasound scattering and acoustic attenuation. The resulting high-volume fraction liquid-filled elastomeric composite exhibits mechanical softness comparable to that of human skin (45 Shore 00), acoustic properties of soft tissues, and excellent environmental stability under both prolonged ambient exposure and dry and humid conditions. Furthermore, the formulation strategy enables the independent tuning of mechanical compliance and acoustic transmission by separately controlling the elastomer network and the liquid inclusions, thereby providing a versatile materials design framework for application-specific optimization.

The developed coupling material was successfully integrated into a functional portable ultrasound imaging device, demonstrating excellent coupling together with good imaging performance. In addition, the material exhibits mechanical conformability to rigid substrates with complex surface topographies while maintaining complete structural recovery after unloading. These characteristics make the material an attractive candidate not only for wearable ultrasound devices and long-term biomedical monitoring, secure ultrasound-based biometric systems, prolonged therapeutic ultrasound applications, but also for industrial non-destructive testing.

Overall, this work establishes a simple, scalable, adaptive (shape and size), and environmentally friendly materials platform for the development of reusable ultrasound couplants. The proposed design strategy provides a practical pathway towards next-generation coupling materials that combine excellent acoustic transmission, mechanical conformability, and durability, thereby addressing an important materials challenge in both biomedical and industrial ultrasound technologies.

# 4. Materials and methods

## 4.1 Materials used

Sylgard 184 is purchased from DOWSIL$^{TM}$. Silicone oils with viscosities of 10 cSt, 350 cSt, and 500 cSt are purchased from Sigma, and a vinyl-functionalized silicone oil (500 *cSt*) was purchased from ABCR chemicals. Glycerol, ethylene glycol, diethylene glycol, and triethylene glycol, with purities ≥ 99 %, are purchased from Sigma.

## 4.2 Emulsion preparation

The emulsions were prepared with Sylgard 184 (and silicone oil) as the continuous phase. The two components of Sylgard 184 are mixed in a 10:1 or 10:2 ratio in a glass beaker (inner diameter 5.5 cm) using a mechanical stirrer at 200 RPM for 5 minutes. The mixture is then placed inside a desiccator at low pressure (-800 *mbar*) for 45-60 minutes to remove the bubbles. Once all the bubbles have disappeared, an appropriate amount of silicone oil is added, mixed again using a mechanical stirrer for 5 minutes, and then placed in a low-pressure environment for 25-40 minutes. Once all the bubbles disappeared, the beaker was attached to a four-blade propeller stirrer with a 5 cm fan diameter. The various liquid phases were then added to the oil phase through a tube at a constant rate of 25 $mL\ h^{-1}$ using a syringe pump (Harvard Apparatus, PHD 2000 Infusion) under constant stirring. The temperature was maintained at 20 °C throughout the preparation process using a temperature-controlled water bath.

## 4.3 Curing and storage

The composite pads were prepared by compressing the viscoelastic emulsion between two glass slides with 3D-printed spacers. The customized spacer was attached to the surface of one glass slide. The as-prepared emulsion was then transferred to the mold and placed in a desiccator under low pressure (-800 *mbar*) to remove air bubbles. Once the bubbles were removed, the mold was taken out of the desiccator, and the top glass slide is placed and secured with clips. The mold was then placed inside an oven for curing, i.e., crosslinking. The curing process takes place in three steps: a) 66 ± 1 hrs at 30 $^{o}C$, and b) 2.5 ± 0.2 hrs at 60 $^{o}C$. Afterward, the samples were stored at room temperature for 48 ± 2 *hrs* to complete the crosslinking process. Slow curing at low temperature was crucial for silicone oil extended silicone elastomer, as rapid curing may cause inhomogeneous crosslinking or even defects and cracks[37]. Afterward, the pads were removed from the molds and stored in a closed plastic petri dish.

## 4.4 Microstructure characterization

The pads were cut using a sharp blade, and the cut faces were carefully wiped out. They were placed under low pressure for 12 hours (6 + 6), and the cut surface was wiped again. The cut faces were then exposed to an electron beam for imaging. The measurements were performed using a Thermo Fisher Quattro SEM under high-vacuum conditions ($10^{-3}$ *Pa*) with an applied voltage in the range of 7-20 *kV*.

### 4.5 Acoustic testing

A dedicated acoustic characterization setup has been established (shown in **figure S8**). The key component of the setup is a couple of PVDF ultrasonic transducers (diameter 10 mm) purchased from Precision Acoustics, UK. The central frequencies are 11.12 MHz and 10.88 MHz with bandwidths of 8.24 MHz and 8.38 MHz (-3 dB), respectively. A high-frequency signal was generated using a 5-level pulsar acquired from Texas Instruments (TX7316EVM). The setup works in transmission mode. Both the ultrasound generator and receiver are fixed on the ends of a customized 3D-printed well. The well is filled with deionized water, and measurements are taken by dipping the pads into the water between the two transducers. The measurements were carried out over a frequency range of 7.7-14.3 MHz. The cured pads were dipped into the measurement well at three positions between the generator and receiver for two different portions of the pad. 5 measurements were taken, and their mean and standard deviation are used for further analysis. The acquired acoustic signals were processed using a Python script optimized for the specific data output. A few examples of data output are listed in **figure S8** in the supplementary information.

### 4.6 Mechanical testing

The compression tests were performed using a rheometer (TA instrument, AR G2) equipped with a Peltier base. A cylindrical specimen with a diameter of 10 ± 0.1 *mm* and a thickness of 4 ± 0.2 *mm* (diameter-to-thickness ratio ≥ 2.5) is used [44],[45]. The compression tests are conducted using a parallel-plate geometry with a diameter of 20 *mm* and an axial compression rate of 10 $\mu m\ s^{-1}$. The disk-shaped sample was cut from the cured pad, placed between the plates, and compressed at a constant rate till 10-15 % of the initial sample thickness, and the recorded axial force and gap data were converted to a stress-strain curve. Young's modulus was calculated from the slope of the stress-strain curve in the strain range of 1 to 10 %. For each sample, 3 measurements were taken, and the mean and standard deviation are used for further analysis. The repeatability of measurements was ensured by measuring pads of different batches and over time.

Further, the hardness of the samples was measured using a durometer. For these measurements, a large sample with a diameter of 30 mm and a thickness of 12 *mm* was used.

### 4.7 Mechanical conformability test

These tests were performed using a 3D printed stiff object with wavy surface features. The stiff object was placed on the bottom plate of a rheometer. The couplant was then placed on top of the wavy pattern of the object. The top plate of the rheometer was then brought into contact with the pad. A white light source was placed on the left side of the rheometer stage, and a digital camera was positioned colinearly with the wavy surface of the object on the right. The top plate of the rheometer was then lowered until the couplant conformed to the stiff target, then retracted. The process was repeated multiple times to check reversibility.

### 4.8 Ultrasound imaging

The soft tissue-mimicking phantom was prepared by mixing deionized water, glycerol, and agarose powder in an 85:13:2 weight ratio. The mixture was heated to 90 *°C* to dissolve the agarose. The hot mixture was poured into a custom-made mold and allowed to cool down to room temperature for gelification. The sound speed in the phantom is ~1520 $ms^{-1}$. These phantoms were used for ultrasound imaging experiments using EcrinLab.

The EcrinLab provides a versatile platform for interfacing a wide range of ultrasound sensors. The EcrinLab platform offers a wide range of adjustable operating parameters, including excitation frequency (5 - 15 *MHz*), transmission voltage ($T_X$ HV), receiver analog frontend (AFE) gain, sound speed of the propagation medium, and dynamic optimization of multiple imaging parameters depending on the selected imaging mode. The system provides excellent imaging performance, enabling volumetric reconstruction with a typical voxel size of approximately half of the ultrasound wavelength (~70 *μm* at 10 *MHz*), together with low noise and high frame rates. For static scenes, the device supports B-mode imaging, full 3D volume reconstruction, and surface imaging, thereby suitable for deep tissue imaging, biometrics, Doppler ultrasound, and prolonged experiments.

## 5. Supporting information

All the supporting information is provided in the supplementary information.

## 6. Acknowledgments

The authors acknowledge Prof. Stéphane Holé, ESPCI Paris, for facilitating the stability test of the couplant under controlled temperature and relative humidity. S.R. acknowledges the use of ChatGPT (GPT 5.5) to visually refine and generate selected schematic elements used in figure 1 based on original schematics prepared by the author. The AI-assisted elements were reviewed and subsequently assembled by S.R. in Microsoft PowerPoint to produce the final version of figures.

## 7. Conflict of interest

All the authors are co-inventors of a patent filed by ID4US, ESPCI Paris, and CNRS (French unexamined patent applications FR2603474, filed on 30/03/2026) related to this work.

## 8. Author contributions

**S.R.**: Conceptualization, Methodology, Data curation, Investigation, Formal analysis, Visualization, Validation, Writing - original draft, Writing - review & editing

**R.F.**: Instrumentation, Software, Data curation, Investigation, Writing - review & editing

**J.B.**: Supervision, Validation, Visualization, Writing - review & editing

**N.B.**: Conceptualization, Methodology, Validation, Visualization, Supervision, Writing - review & editing, Project administration, Funding acquisition

# Supplementary information

## Tailoring Mechanical and Acoustic Properties of Liquid-Filled Elastomers as Reusable Ultrasound Couplants

Surojit Ranoo[1*], Romain Fayolle[2], Jean Baudry[1], Nicolas Bremond[1*]

1. LCMD, CBI, ESPCI Paris, Université PSL, CNRS 75005 Paris, France
2. ID4US, Saint Martin d'Heres, FRANCE

Correspondence: Surojit Ranoo (surojit.ranoo@espci.fr) and Nicolas Bremond (nicolas.bremond@espci.fr)

## S1. Materials formulation

Emulsions were prepared using sylgard 184 as the continuous phase. The silicone elastomer consists of two components (part A and part B) with a recommended mixing ratio of 10:1 (A:B) and exhibits a viscosity of approximately 3.5 *Pa s*. The preparation of high internal phase emulsions $\varphi > 70$ % with this high viscous continuous phase without surfactants remains challenging. Recently, Nannette *et al.* demonstrated that stable emulsions containing up to approximately 80 vol. % of dispersed phase can be obtained by closely matching the viscosities of the silicone oil and aqueous liquids[1]. In the present system, the viscosity of the silicone elastomer is approximately twice that of glycerol. To improve viscosity matching, silicone oils with different PDMS chain lengths (viscosities) were incorporated into the elastomer mixture to reduce the effective viscosity of the continuous phase. The silicone oils are chemically compatible with the silicone elastomer and remain inert during the crosslinking reaction, acting primarily as plasticizers that soften the elastomer matrix. Based on both processability and mechanical performance, an optimum formulation was obtained with 30 % silicone oil with a viscosity of 350 *cSt*. The resultant viscosity of the continuous phase was 1.12 *Pa s*.

Using this modified silicone elastomer, an emulsion formulation was possible up to 75 % using glycerol as the dispersed phase. The stability of the surfactant-free emulsions prepared with high-viscosity PDMS arises from the formation of thin interdroplet films that become trapped between neighbouring droplets, producing a strongly adhesive and viscoelastic system. Achieving this stability required a precise balance between the viscosities of the continuous (PDMS) and dispersed liquid phases. For glycerol/silicon elastomer emulsions with comparable viscosities, the emulsion became highly viscoelastic when $\varphi$ exceeded *68 %* , making further incorporation of the dispersed phase increasingly difficult. Above 70 %, the propeller-type stirrer was no longer sufficient to homogeneously incorporate the glycerol. Therefore, after initial emulsion preparation, the emulsion was transferred to a mortar and mixed with a pestle. This additional processing significantly improved the homogeneity of the emulsion while simultaneously reducing the droplet size.

In contrast, because of large viscosity mismatch between the continuous and dispersed phases, stable emulsions with $\varphi$ exceeding 64 % could not be prepared using EG, DEG, and TEG. These systems underwent phase separation close to random close packing, preventing the formulation of a stable emulsion with a high $\varphi$.

After the preparation, the emulsions were cast on a custom-designed mould and cured under controlled conditions. The microstructure of the cured composites was investigated using electron microscopy. The compositions with glycerol exhibit a closed cell structure (**figure S2**).

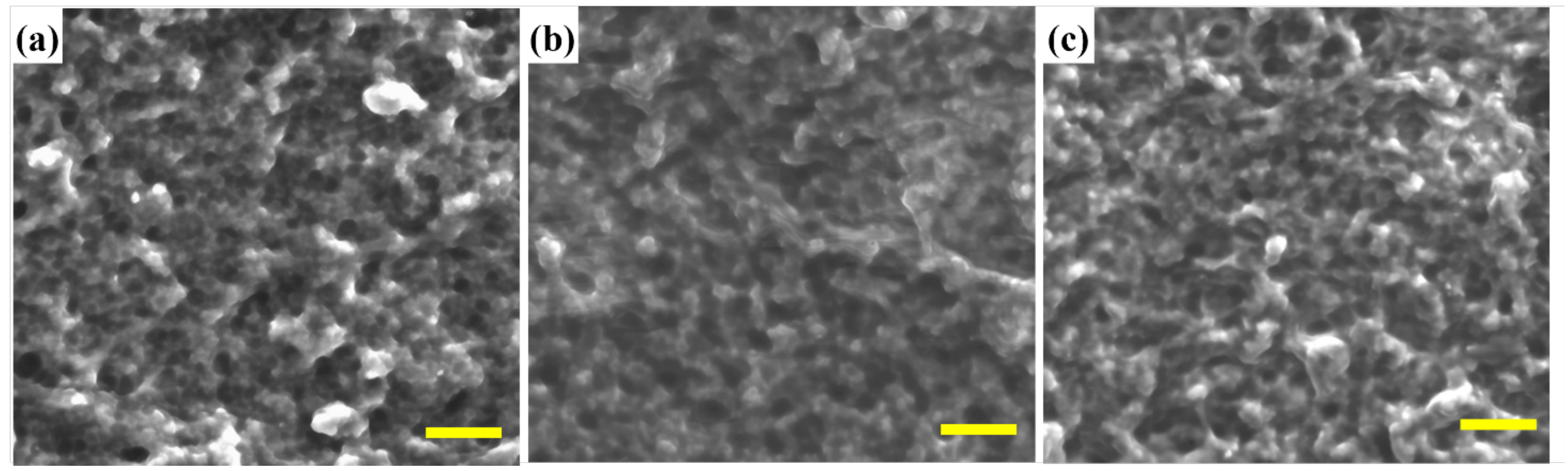

**Figure S1**: Microstructure of solid composites incorporated with glycerol volume fractions of (a) 50 %, (b) 60 %, and (c) 70 %. The scale bar corresponds to 10 $\mu m$.

## S2. Evolution of droplet size, adhesion, and viscoelasticity of emulsions

Optical microscopy images of the as-prepared emulsions are presented in **figure S2**. The droplet size and size polydispersity increased slightly with increasing glycerol volume fraction. These emulsions were highly stable and strongly adhesive owing to the formation of a thin, stable interdroplet PDMS film[1]. This adhesion behavior was particularly evident in microscopy images of emulsions diluted 20 times in hexadecane. The thin PDMS film remained partially intact even after dilution, confirming the robustness of the emulsion structure (**figure S3**). As shown in **figure S2**, below the random packing limit ($\varphi < 64\%$), the droplets remain relatively small even after dilution, indicating good structural stability. However, at higher inclusion levels ($\varphi \sim 70$-$75\%$), the droplets became tightly packed and deformed due to the higher contact forces between neighboring droplets. These emulsions tend to coalesce faster after dilution.

Emulsions containing more than 50 % of glycerol exhibited viscoelastic behavior together with well defined yield stress, as shown in the **figure S4a**. The yield stress increased one order of magnitude as $\varphi$ increased from 50 to 70 %. The high yield stress value of approximately 500 Pa for $\varphi = 70\%$ is attributed to the strong droplet adhesion and the formation of a droplet network. This behavior is reflected in the high viscosity of emulsions at low shear (**figure S4b**). Nevertheless, the emulsions showed pronounced shear-thinning behavior, which facilitated processing despite their high viscoelasticity.

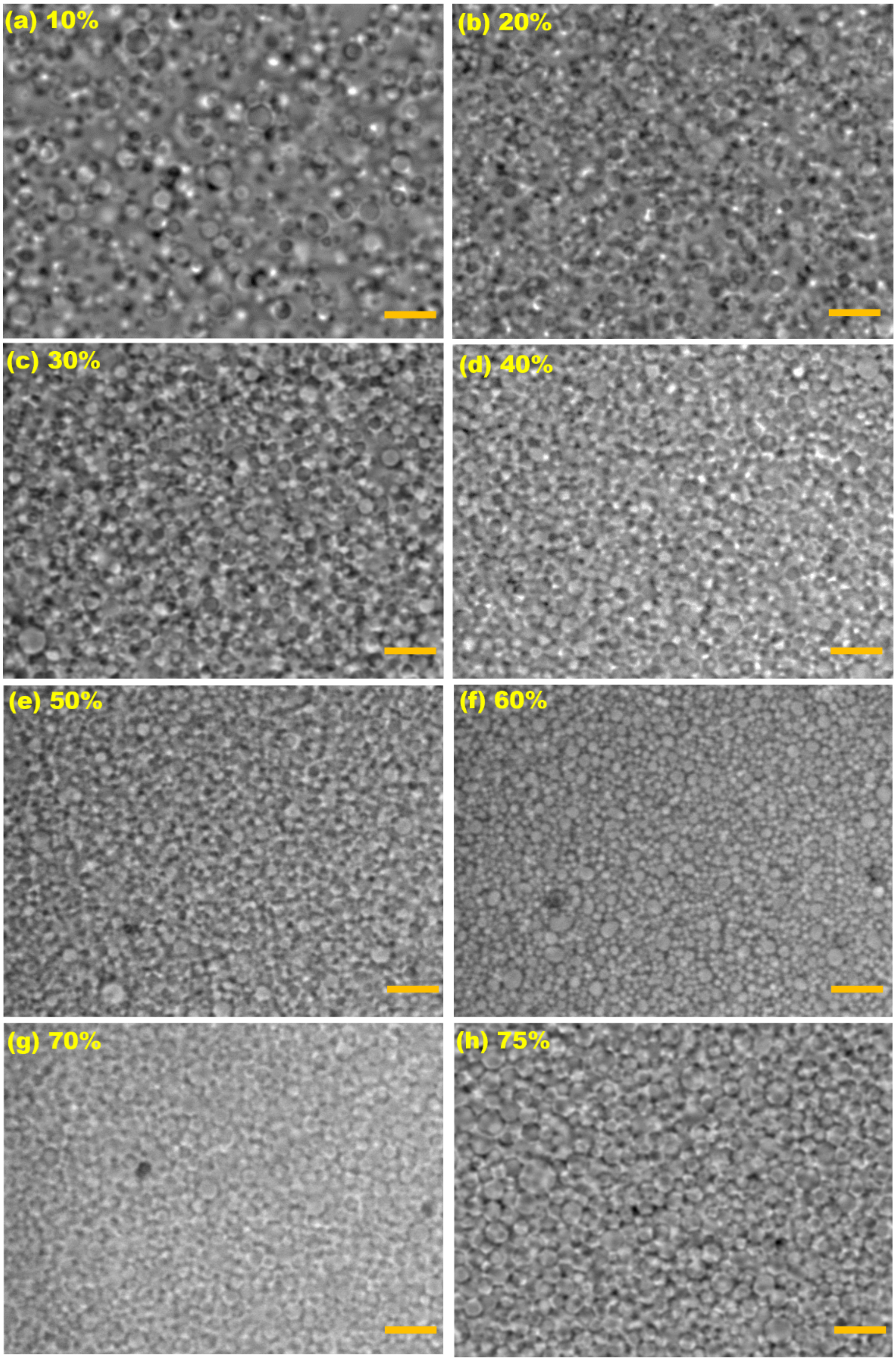

**Figure S2**. Optical microscopy images of as-prepared emulsion containing various volume fractions of glycerol. The scale corresponds to 10 $\mu m$.

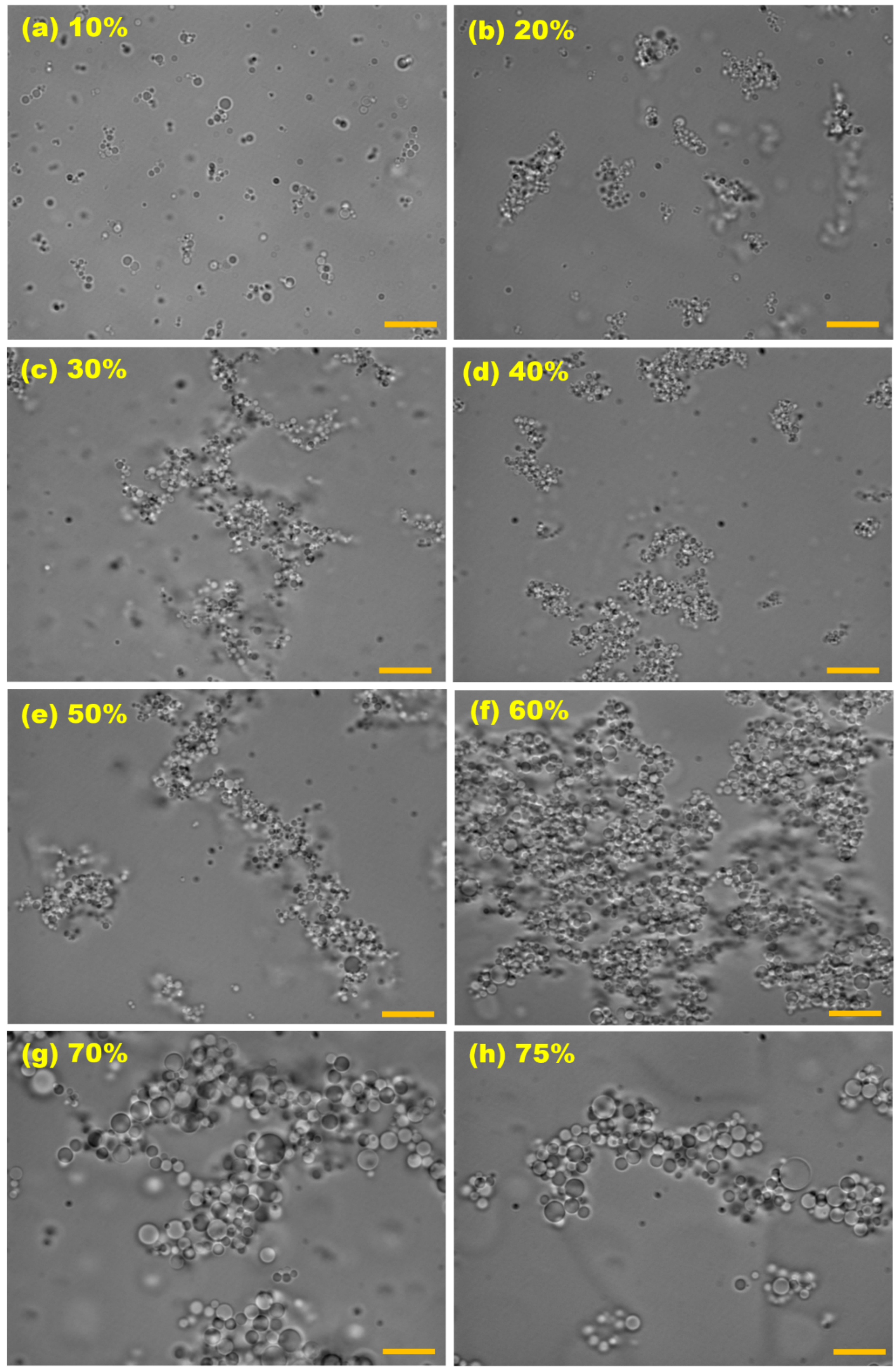

**Figure S3**: Optical microscopy images of diluted emulsions containing various vol. fractions of glycerol. The as-prepared emulsions were diluted 20 times in hexadecane. The scale corresponds to 10 $\mu m$.

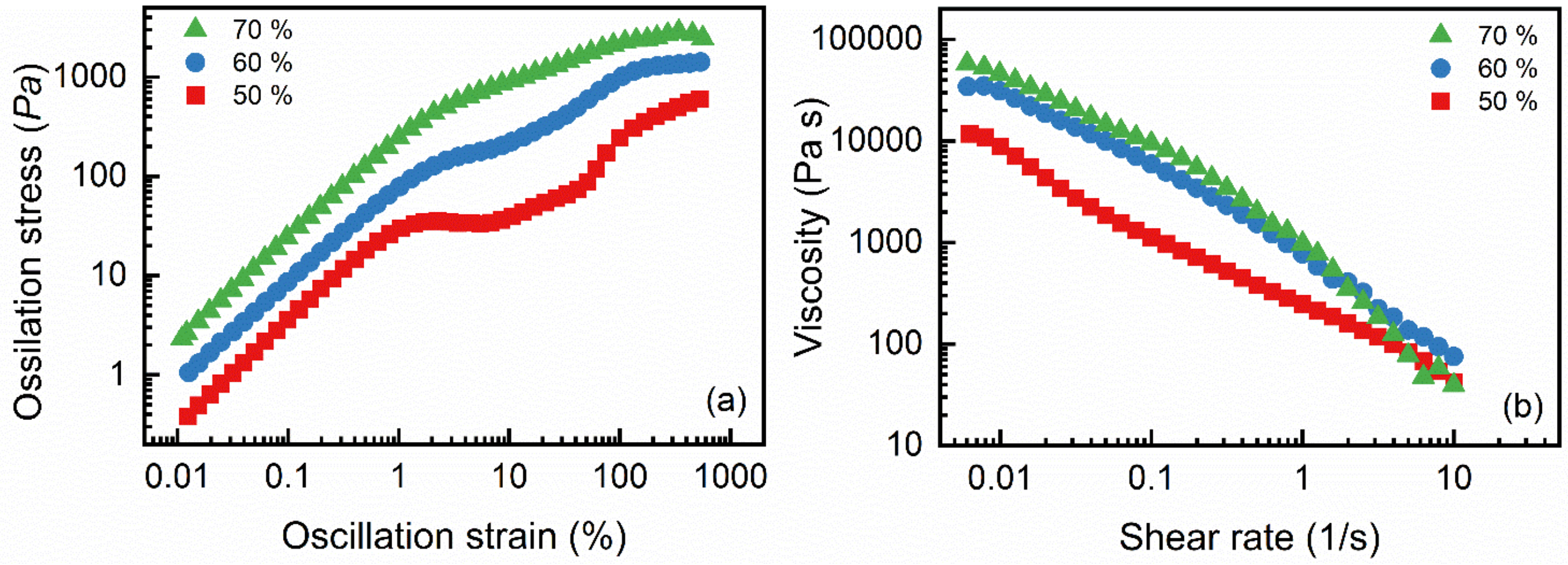

**Figure S4:** (a) Amplitude sweep test for as-prepared emulsion with 50 %, 60%, and 70 % of dispersed phase (glycerol). (b) Flow curve of the emulsion with 50 %, 60%, and 70 % of dispersed phase.

## S3. Consistency and reproducibility of measurements

The reproducibility and consistency of the measurements were verified before finalizing the measurement protocols. **Figure S5** demonstrates the excellent repeatability of the compression test over 15 consecutive compressions for composites formulated with glycerol. The reproducibility of material properties was further validated by testing samples prepared from different batches and characterized at different time intervals. As shown in **figure** S6, both the acoustic and mechanical properties remained consistent, confirming the robustness of the formulation and fabrication process. Further, the long-term stability of the developed pads was subsequently evaluated by storing the samples in an open petri dish in a laboratory environment (24 ± 3 ℃) for a period of 5 months (March to July). The mass of the pad was monitored initially after 2 weeks and subsequently after 4 weeks interval. Over the 6 months period, the pads exhibited a change in mass of 4.8 %, indicating excellent stability under ambient conditions. The speed of sound measured after 6 months decreased 1 %. To further evaluate the stability, additional tests were performed under controlled conditions at 30 ± 0.1 ℃ and 30% relative humidity for 15 days, followed by 30 ± 0.1 ℃ and 40% relative humidity for 7 days. Under these conditions, the pad exhibited a change in mass of approximately 1.4 % and a 2.1 % decrease in sound speed after 21 days.

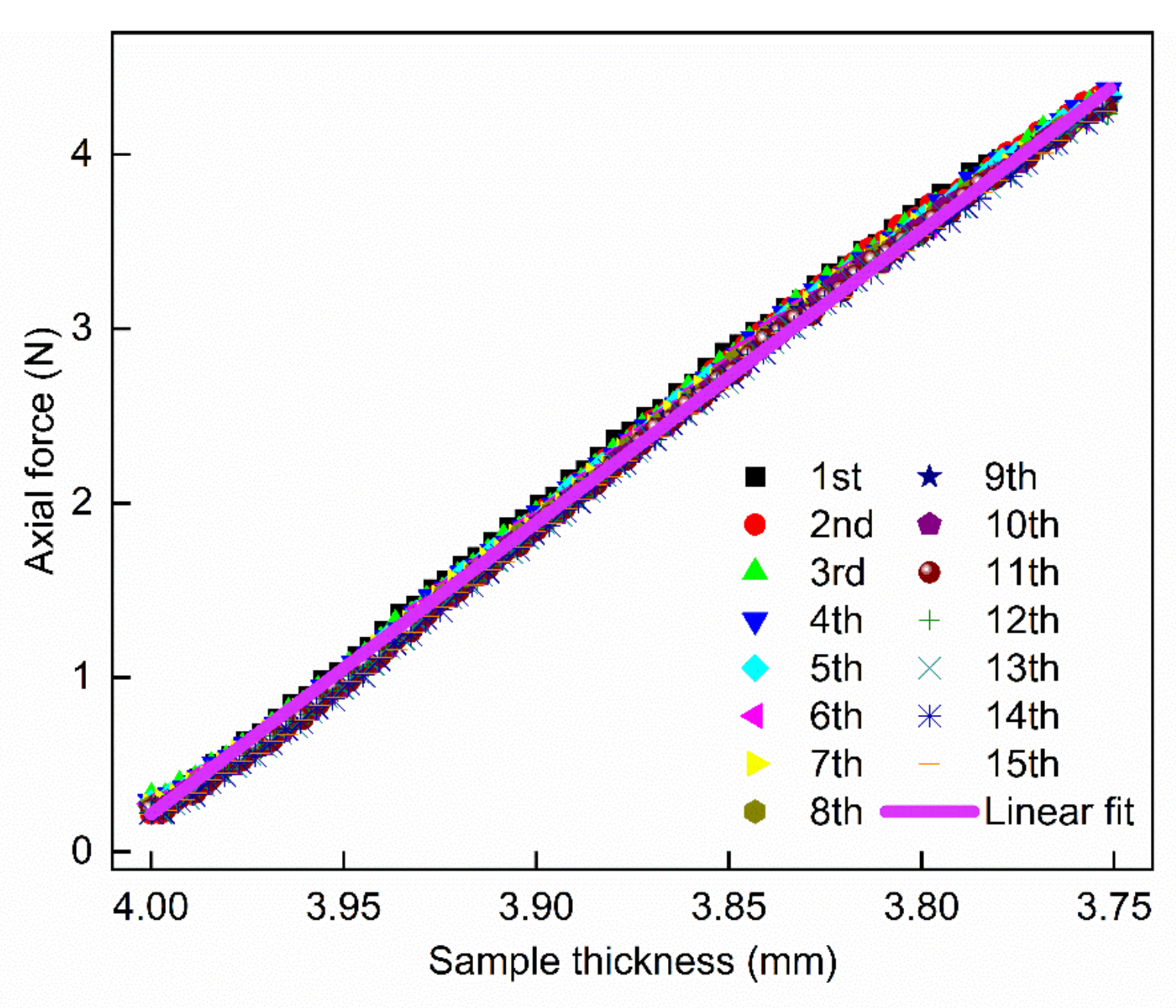

**Figure S5**. Consecutive compression test measurements for composite pads with 70 % of glycerol inclusion.

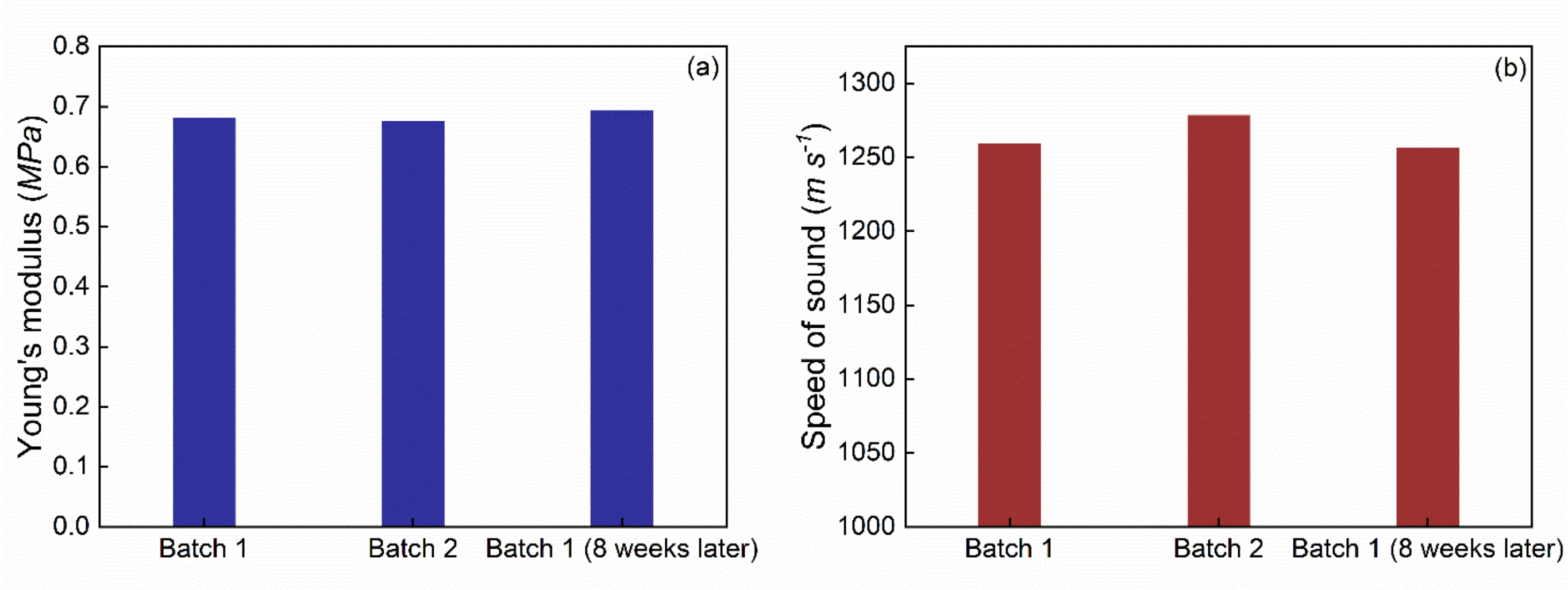

**Figure S6**. Measurements of material properties of pads (glycerol 50%) over time for different batches. Variation of (a) Young's modulus, and (b) speed of sound.

## S4. Evolution of droplet size for emulsions with different dispersed phases

Emulsions were prepared using different dispersed phases to examine the influence of liquid composition on emulsion properties. **Figure S7** shows the microscopy images for emulsions containing 50 vol. % dispersed phases. The droplet size was significantly larger for emulsions prepared with diols than for those prepared with glycerol. This behavior primarily arises from viscosity mismatch between the continuous and dispersed phases, as the viscosities of EG, DEG, and TEG are approximately one order of magnitude lower than that of glycerol. Interestingly, the largest droplet size was observed for TEG, despite it being the most viscous among the glycols. This observation suggests that droplet formation is governed not only by the viscous effect but also by additional interfacial effects.

It has been reported that hydroxyl (*-OH*) groups of glycols can interact with components of the sylgard 184 precursor (Part A), leading to the formation of amphiphilic species that modify the interfacial properties[2]. This effect becomes increasingly pronounced for long-chain glycols. Ginot *et al.* demonstrated the formation of such amphiphilic structures in systems where long-chain glycols are grafted onto methylhydrosiloxane-dimethylsiloxane copolymers (MHRD, trimethylsiloxy-terminated)[3]. The resulting molecular architecture closely resembles that of commercial silicone emulsifiers, implying a similar surfactant-like behavior at the droplet interface[4]. Such interfacial interactions are expected to influence the stability of the thin inter-droplet PDMS film and, consequently, the emulsion microstructure.

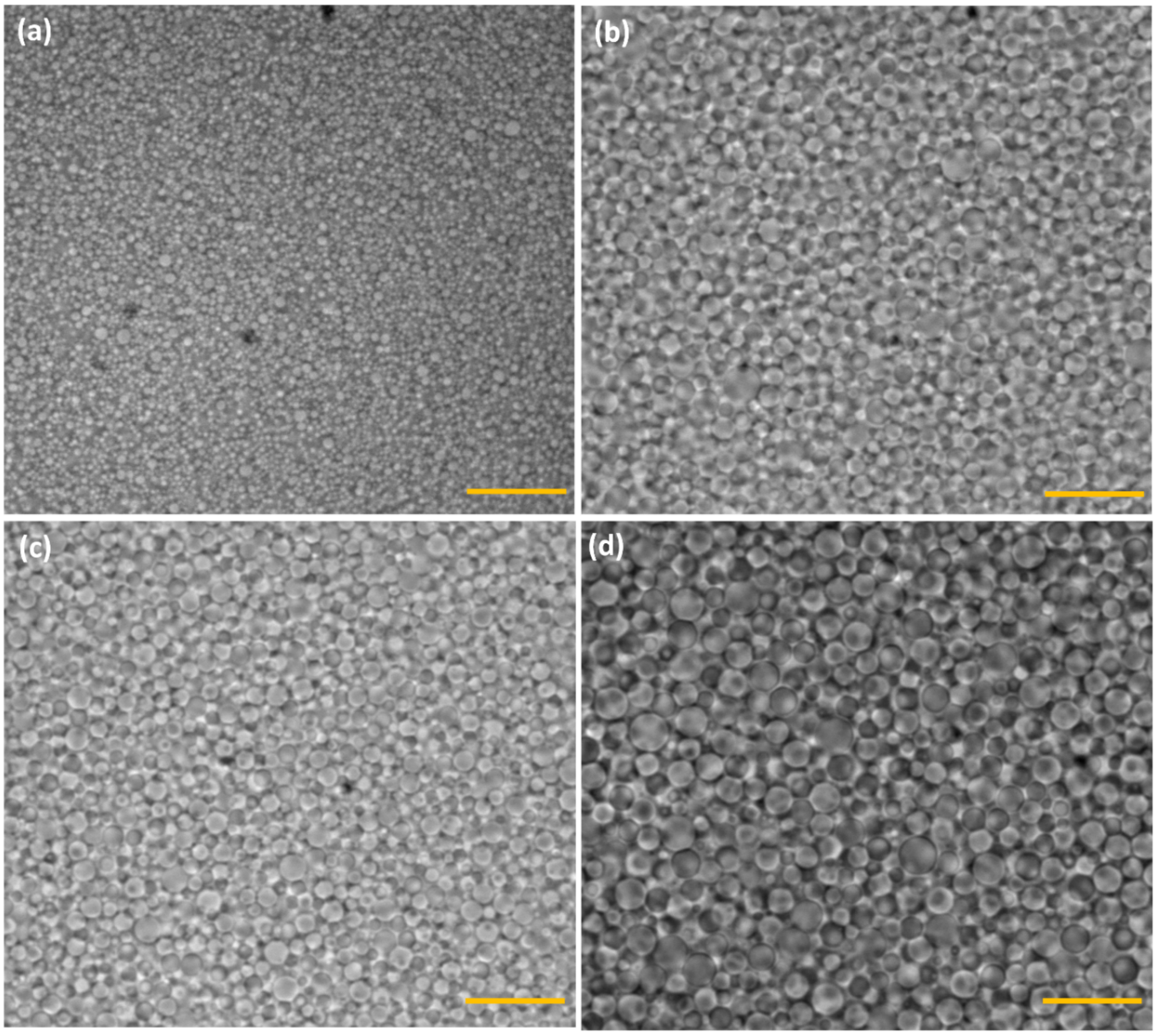

**Figure S7**. Optical microscopy images of as-prepared emulsions (vol. 50%) with different dispersed phases: (a) glycerol, (b) ethylene glycol, (c) diethylene glycol, and (d) triethylene glycol. The scale bar is 10 μm.

## S5. Acoustic characterization setup

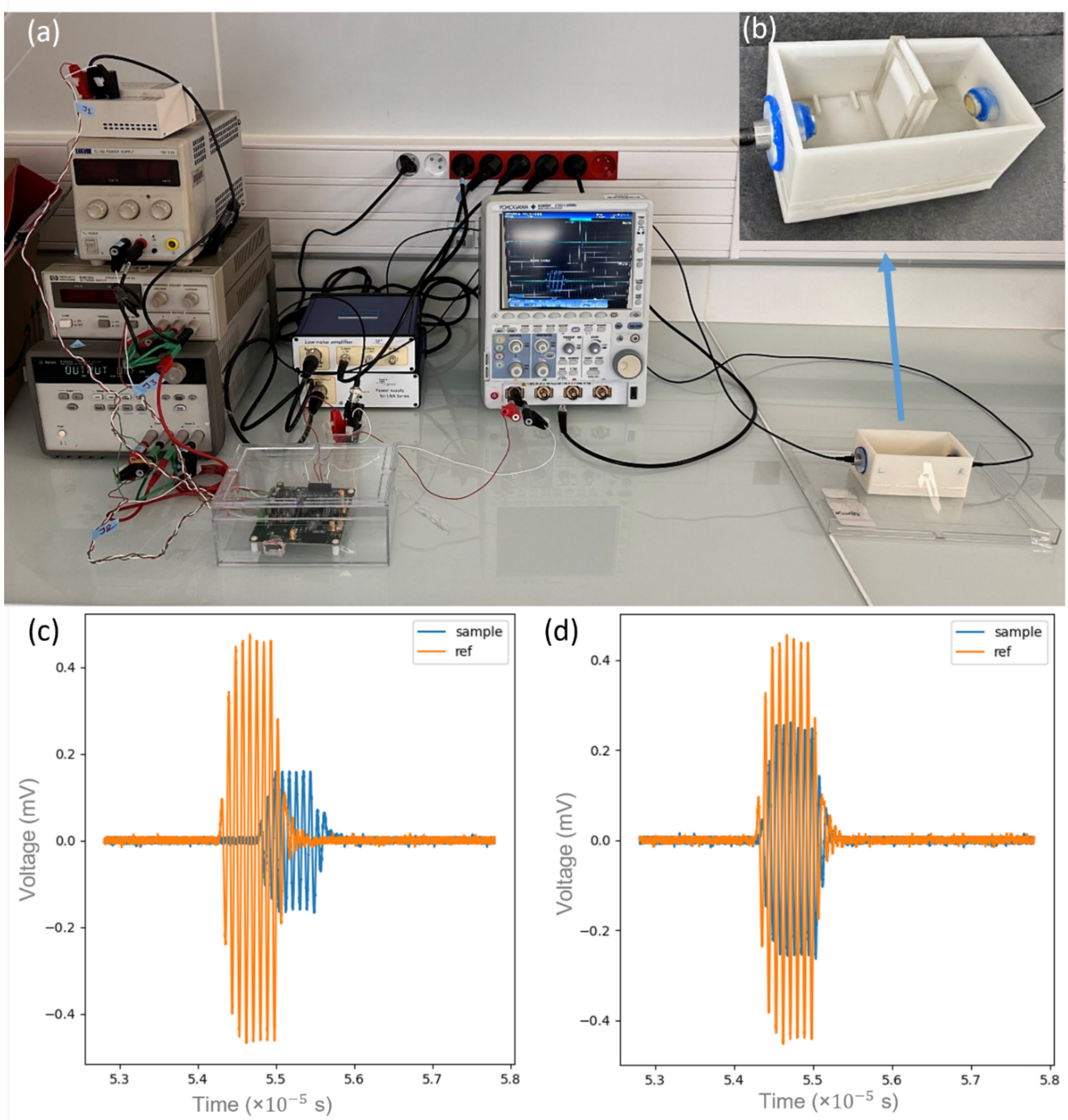

**Figure S8**. (a) A custom-made acoustic characterization setup for the estimation of sound speed in the time-of-flight method. (b) A water-filled chamber with 2 ultrasound transducers. Ultrasound pluses propagate through the reference medium (water) and (c) through pure elastomer pads and (d) through a composite pad with 75 % glycerol inclusion.

## S6. Acoustic impedance

The propagation of an ultrasound wave across an interface between two media is governed by the acoustic impedance of the media. The intensity transmission coefficient is defined as $T = 4z_1z_2/(z_1 + z_2)^2$, where $z_1$and $z_1$are the impedances of the media 1 and 2, respectively. Efficient ultrasound transmission, therefore, requires acoustic impedance matching between the couplant and targets. The acoustic impedance is calculated considering the speed of sound ($v_c$) and density ($\rho$) as $z = \rho v_c$. **Figure S9** shows the variation of $z$ of composites with glycerol volume fraction. Similar to the speed of sound, $z$ increases with glycerol content, as speed of sound and density both are volumetric properties. The family of liquid-filled elastomeric composites spans the acoustic impedance range of several biological tissues, including fat, skin, blood, and muscle, demonstrating the ability to achieve effective acoustic impedance matching for ultrasound applications.[5]

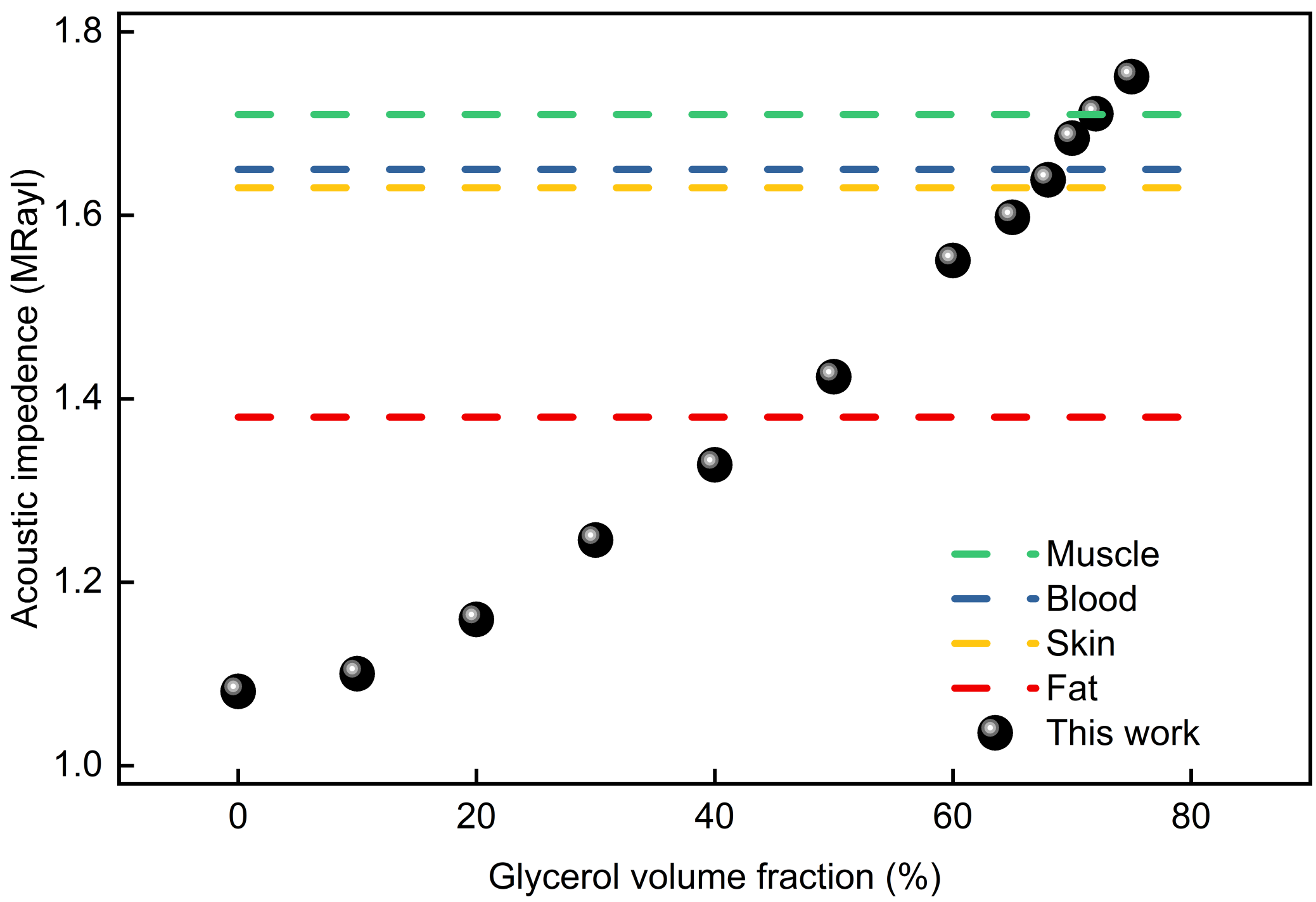

**Figure S9**. Variation of acoustic impedance of the composite with increasing glycerol volume fraction and indication of acoustic impedance of selected human body tissues.